\documentclass[12pt]{article}

\usepackage{amsmath}
\usepackage{amssymb}
\usepackage{rotate}
\usepackage{graphics}
\usepackage{graphicx}
\usepackage[sort&compress,square,comma,numbers]{natbib}
\usepackage{seceqn}
\usepackage[a4paper]{geometry}

\newcommand{\celsius}{^\circ}
\usepackage{soul}
\usepackage[colorlinks=true]{hyperref}  
\usepackage{calrsfs}

\date{}

\begin{document}

\title{Impact of energy dissipation on interface shapes and on rates for dewetting from liquid substrates\\[1cm]}

\author{
Dirk Peschka\thanks{E-mail: dirk.peschka@wias-berlin.de} $^{1}$,
\and \hspace*{-1.0cm}
Stefan Bommer$^2$, 
\and \hspace*{-1.0cm}
Sebastian Jachalski$^{1}$, \\
\and \hspace*{-1.0cm}
Ralf Seemann$^2$, Barbara Wagner$^{1}$
\\
\small{$^1$ Weierstra{\ss} Institute, Mohrenstrasse 39, 10117 Berlin, Germany} \\
\small{$^2$ Experimental Physics, Saarland University, 66123 Saarbr\"ucken, Germany}\\
} 

\maketitle

\vspace{1cm}

\allowdisplaybreaks

\begin{abstract}
We revisit the fundamental problem of liquid-liquid dewetting and perform a detailed comparison of theoretical predictions based on thin-film models with experimental measurements obtained by atomic force microscopy (AFM).
Specifically, we consider the dewetting of a liquid polystyrene (PS) layer from a liquid polymethyl methacrylate (PMMA) layer, where the thicknesses and the viscosities of PS and PMMA layers are similar.
The excellent agreement of experiment and theory reveals that dewetting rates for such systems follow no universal power law, in contrast to dewetting scenarios on solid substrates.
Our new energetic approach allows to assess the physical importance of different contributions to the energy-dissipation mechanism, for which we analyze the local flow fields and the local dissipation rates.
\end{abstract}
%


\section{Introduction}
The evolution of many physical systems is governed by thermodynamical or mechanical energetic principles, \textit{e.g.}, \cite{grmela1997dynamics,morrison1998hamiltonian,de2013non,otto2001geometry}.
Such principles are versatile instruments that allow the derivation of underlying physical equations \cite{peletier2014variational}. For flows of incompressible liquids, energy-dissipation principles are known for a long time \cite{helmholtz1868theorie,korteweg1883xvii}.
In particular for thin-film flows, the great success in the quantitative understanding of viscous
flows with contact-line motion has supplied a universal tool that enables the nano- and microstructuring and functionalization of surfaces, but moreover allows to relate flow patterns with liquid properties and substrate chemistry, \emph{e.g.}, as in \cite{Gentili2012}.
Typical phenomena governed by such principles are the dewetting of liquids from solid substrates and from liquid substrates, or in general for wetting and spreading phenomena \cite{oron1997long,fetzer2005new,CrasterMatar2009,joanny1987wetting,bro}, where the balance of the decline of energy and the dissipation $\dot{\mathcal{E}}=-\mathcal{D}\le 0$ can be used to derive power-law rates for the velocity of moving contact lines.

The exponents of such power-law rates depend on the dominant physical effect, \textit{e.g.},
gravity, surface tension, viscous dissipation in bulk and on interfaces, and the geometry of the problem \cite{bonn2009wetting}.
The basic assumption behind such rate estimates is a simple relationship between the energy and the shape of the domain, often also including assertions about the self-similarity of the evolution.
For instance, for a large class of processes where a liquid $(\ell)$ dewetts from a substrate (s) with a straight contact line, the change of surface energy can be  approximated
\begin{align*}
\dot{\mathcal{E}}=\tfrac{\mathrm{d}}{\mathrm{d}t}\bigl(\gamma_\text{s}|\Gamma_\text{s}|+\gamma_\text{s,$\ell$}|\Gamma_\text{s,$\ell$}|+\gamma_{\ell}|\Gamma_{\ell}|\bigr)\sim S \times \dot{x}_\text{c},
\end{align*}
where
$S={\gamma}_{\rm s}-({\gamma}_{\rm s,\ell}+{\gamma}_\ell)<0$
is the spreading coefficient of the system constructed from the corresponding surface- and interface-tensions $\gamma_\alpha$ of interfaces $\Gamma_\alpha$ with surface area $|\Gamma_\alpha|$ and $\dot{x}_\text{c}$ is the contact line velocity.
However, it is problematic to find a general and similarly simple approximation for the energy-dissipation rate
\begin{align*}
\mathcal{D}(\mathbf{u})=\int_\Omega \boldsymbol{\tau}(\mathbf{u}):\nabla\mathbf{u}\,{\rm d}\Omega,
\end{align*}
since shear stress $\boldsymbol{\tau}$ and flow field $\mathbf{u}$ have
a complicated local structure that depends on fine details of the geometry. Therefore, one requires a deeper understanding of the specific dissipation mechanisms 
in order to understand the dynamics of the corresponding processes.
\vspace*{0.1cm}

In the pioneering works for liquid-liquid dewetting by Joanny \cite{joanny1987wetting} and Brochard-Wyart et al.~\cite{bro} dewetting rates for small equilibrium contact angles and for limiting regimes of the liquid-liquid viscosity ratios were predicted. While both works combine valid hydrodynamical and dissipation arguments to derive expressions for contact line velocities, the impact of non-trivial interface shapes on the flow and dissipation remains unclear.
Since then, many theoretical studies are concerned with the derivation of appropriate thin-film models to study the long-time morphological evolution of the liquid layers. Apart from investigations into  stationary states and how they are approached \cite{craster2006dynamics,bommer2013droplets}, a number of studies focussed on modes of instability in liquid-liquid dewetting using stability analysis and numerical simulations of the thin-film models  \cite{pototsky2004alternative,fish, kriegsmann2003steady, Jachalski2014, Pototsky2016} even with additional surfactants \cite{karapetsas2011surfactant}.

On the experimental side dewetting rates and morphologies for liquid-liquid model systems such as  PS on PMMA  have been investigated systematically by  Krausch et al.~\cite{lam,wan} by varying the heights and viscosities of the liquid layers. Similar experimental studies were performed by Pan et al.~\cite{pan} for further layer viscosities and heights.
However, the shapes observed by Krausch et al. \cite{wan} differ considerably from the empirical predictions used to derive dewetting rates \cite{joanny1987wetting,bro}, which were found to be constant.

To the best of our knowledge fundamental dynamic properties, such as dewetting rates have not been settled up to now. The main reason for this is certainly the absence of theoretical confirmations for the observed shapes of dewetting rims, which then might help to understand the mechanisms behind certain dissipation balances and dewetting rates. Additionally, a quantitative study also requires the key parameters of the experimental system, \emph{i.e.}, surface tensions, viscosities, and layer thicknesses, to be determined sufficiently precise. The focus of this study is thus to supply a quantitative understanding of the dewetting mechanics by detailed comparisons of experimentally obtained rim shapes, their evolution, and their dewetting dynamics with those computed from thin-film equations. Additionally, we examine the underlying mechanisms by discussing flow fields and energy dissipation mechanisms.

\section{Experimental setup and model}
\noindent
As a model system we consider a layer of
viscous liquid polystyrene (PS) ($\ell$) above a viscous liquid substrate consisting of polymethyl methacrylate (PMMA) (s).
Initially the two layers are in a glassy state having constant thickness
$h_\ell(t=0,x,y)=\bar{h}_\ell$ for $x>x_c(t=0)$ and
$h_{\rm s}(t=0,x,y)=\bar{h}_{\rm s}$, and are supported by solid silicon wafers at $z=0$. The dewetting process is then started by heating the materials above the glass transition temperature and monitored by \emph{in situ} atomic force microscopy (AFM).
\begin{figure}[h]
 \centering
 \includegraphics[width=0.77\linewidth]{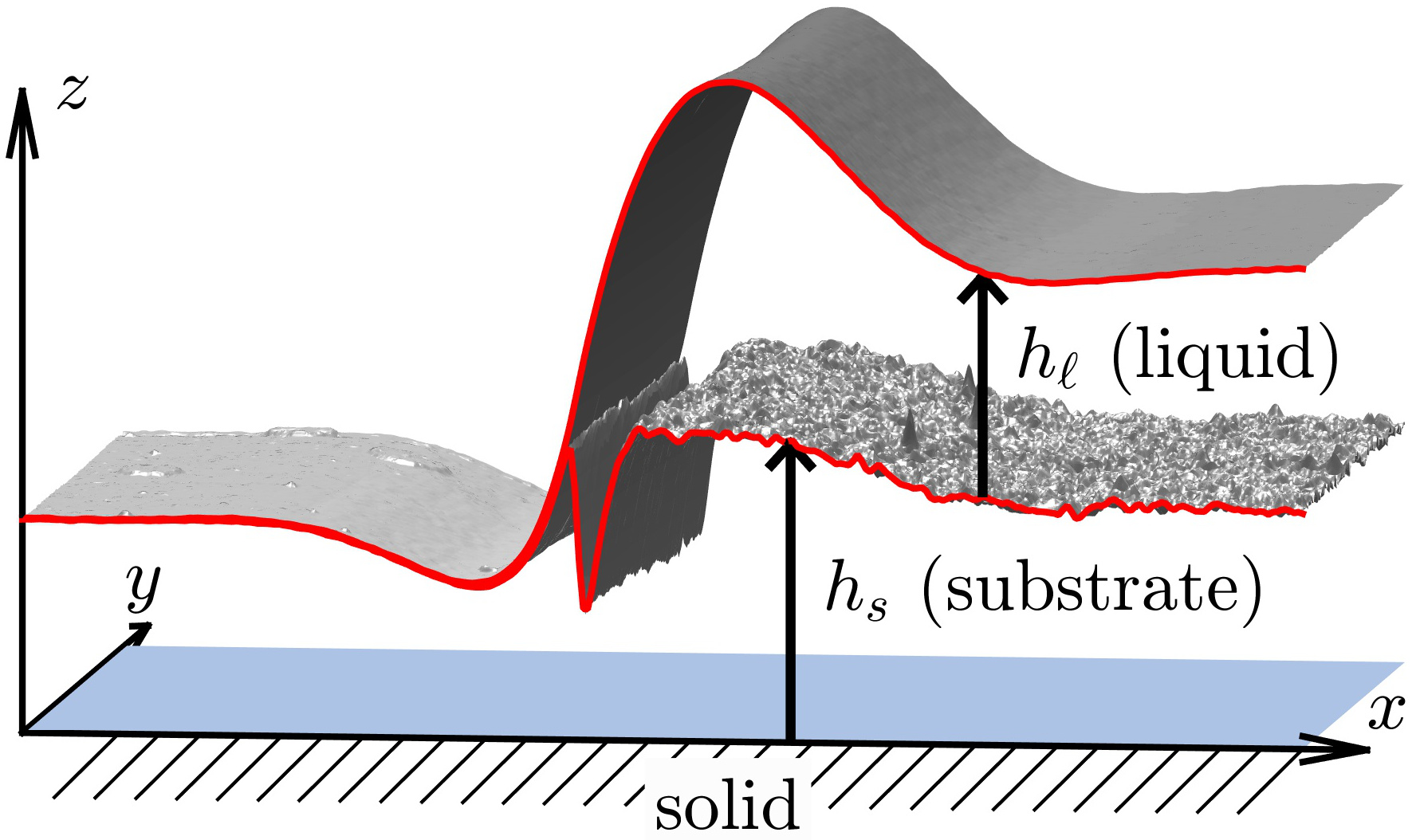}
 \caption{3-$D$ shape of a dewetting rim composed of AFM scans of  PMMA-air, PS-air, and PS-PMMA interface with $\bar{h}_\ell = (125\pm5)\,{\rm nm}$, $\bar{h}_{ \rm s} = (125\pm5)\,{\rm nm}$ after dewetting for $24\,{\rm h}$ at $T=140\,{\celsius}C$.}
 \label{fig:AFM3D}
 \end{figure}


\begin{figure*}[ht]
 	\centering
 	\includegraphics[height=0.217\linewidth] {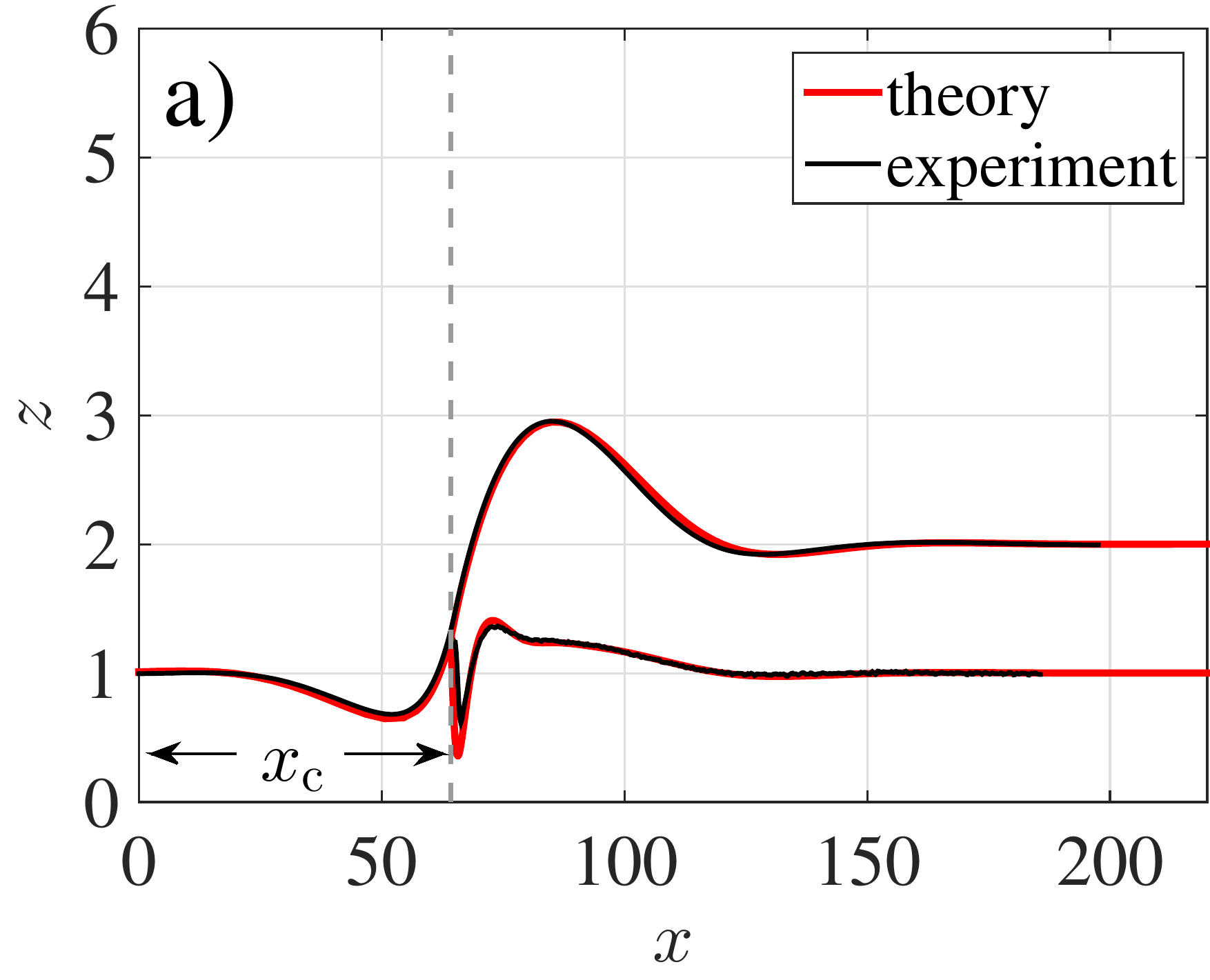}
 	\includegraphics[height=0.2125\linewidth]{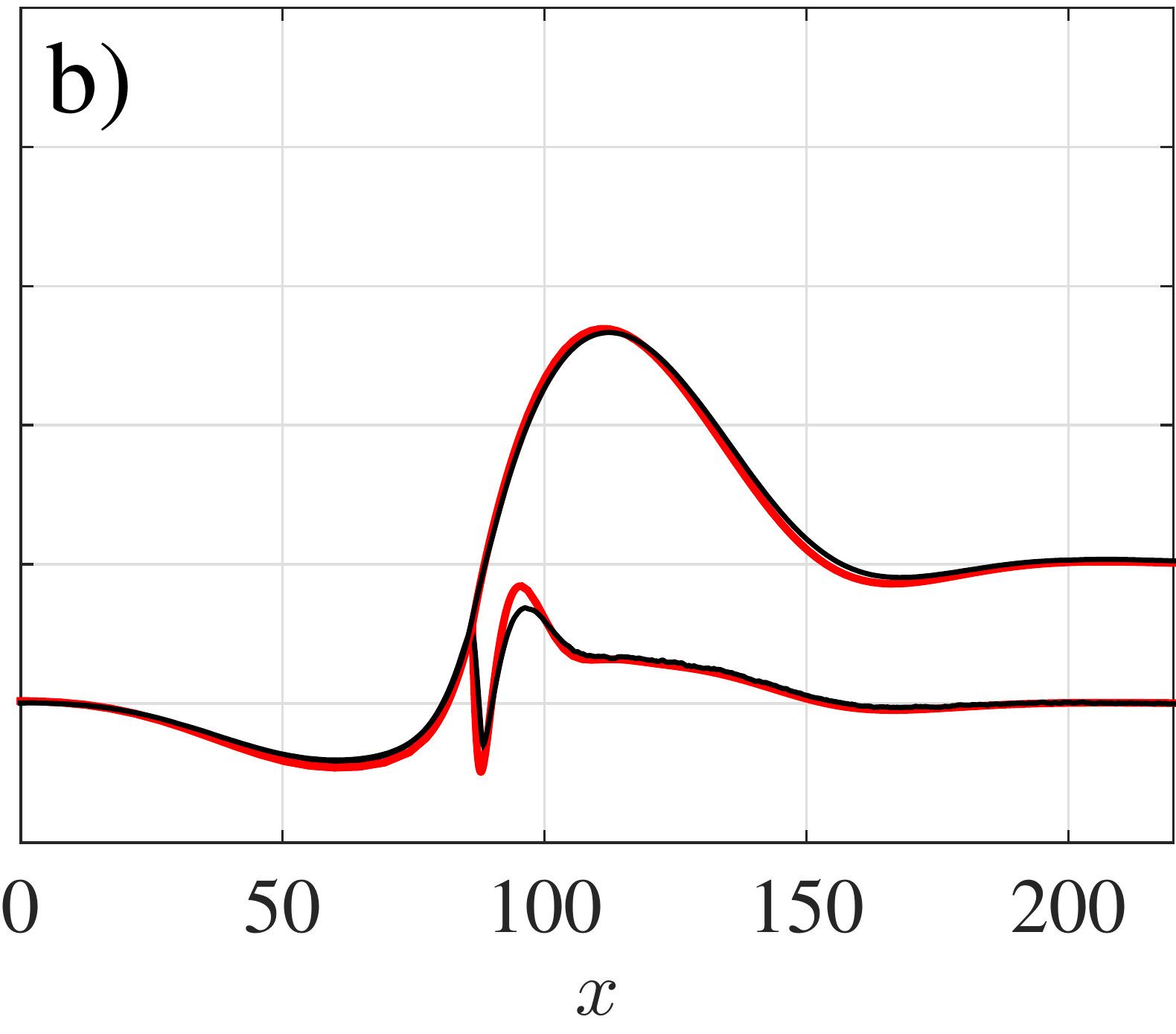}
  \includegraphics[height=0.2135\linewidth]{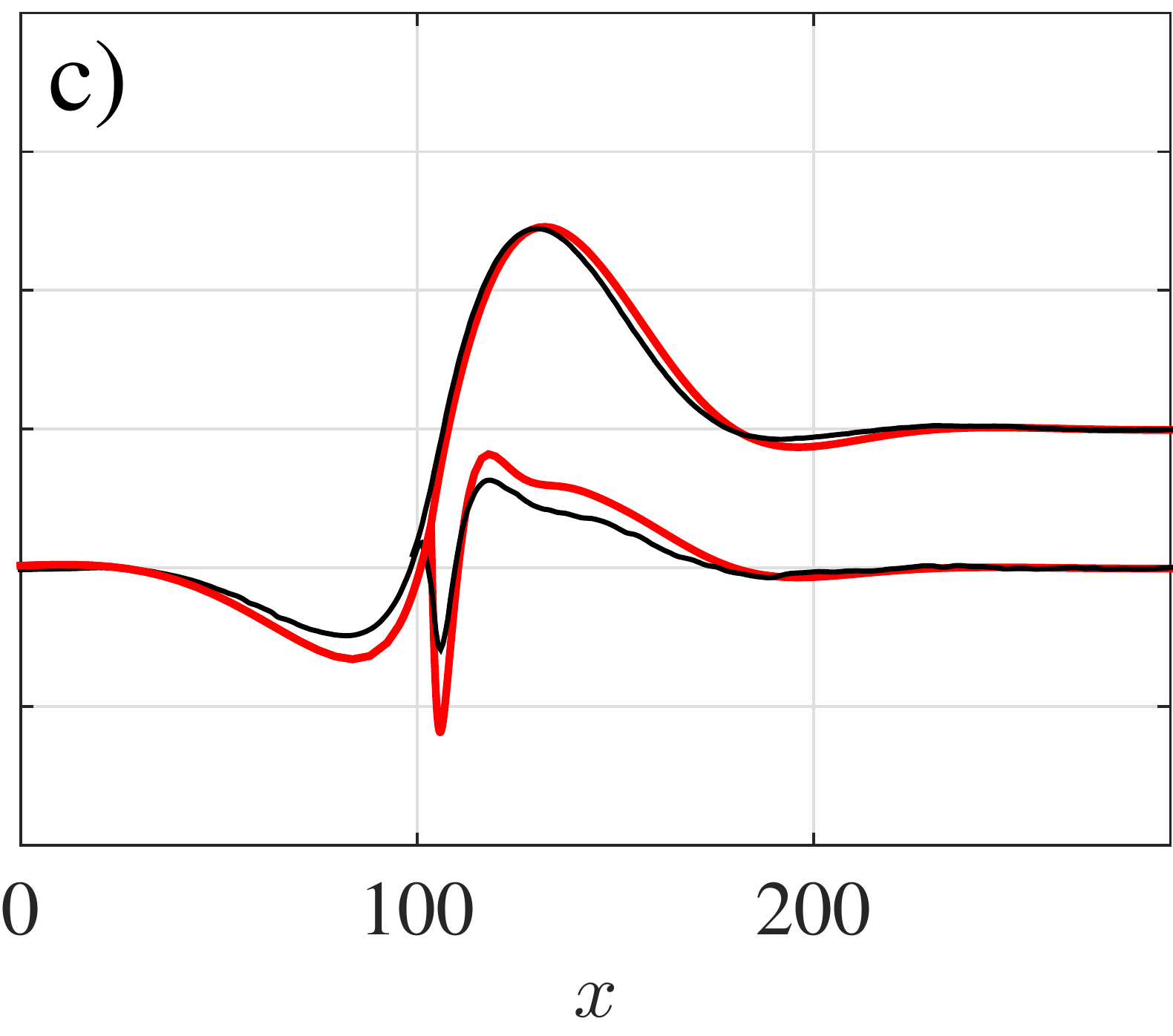}
  \includegraphics[height=0.2135\linewidth]{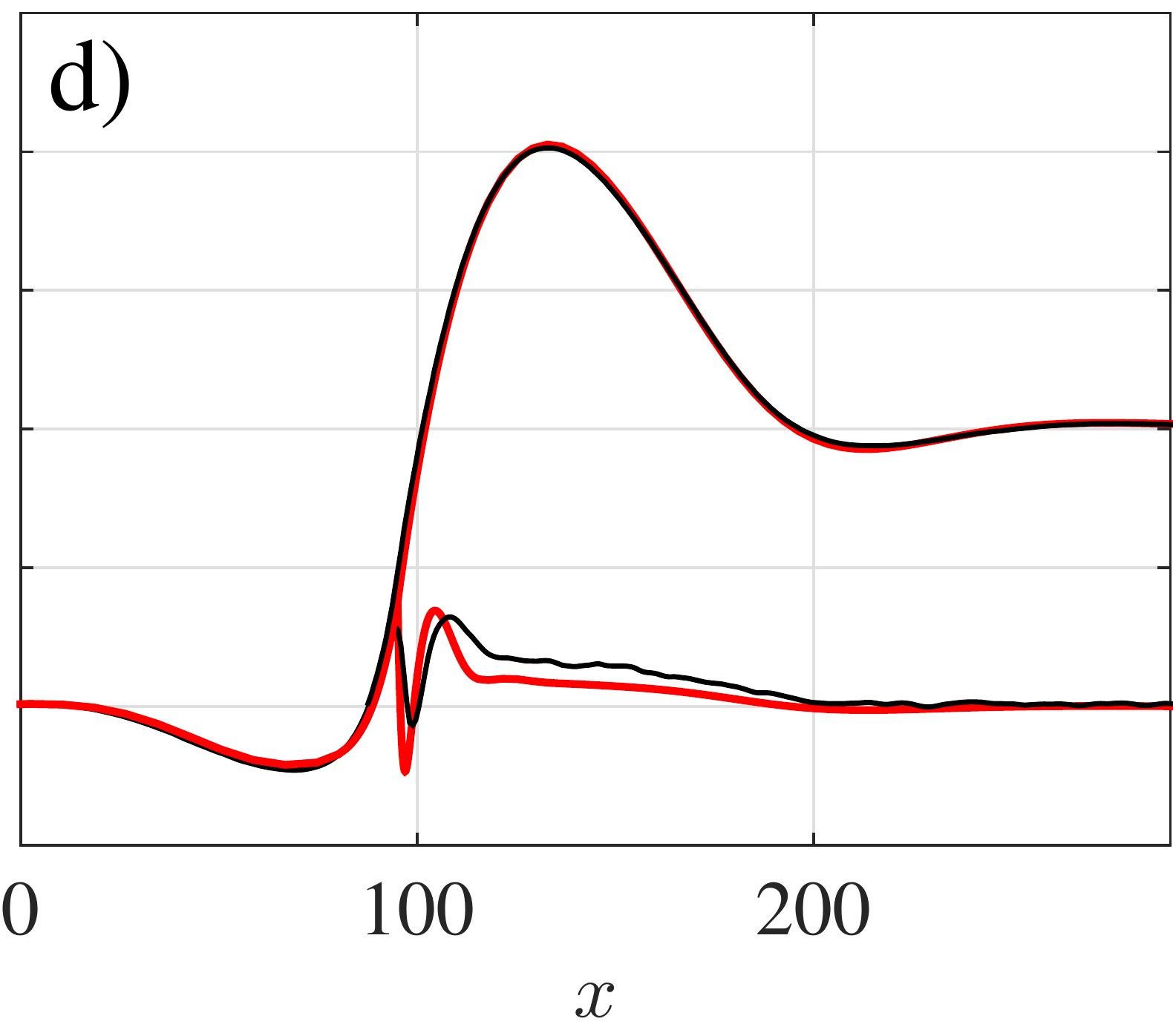}
  \caption{Interfaces from theory and experiment for different thickness ratios $\bar{h}_\ell {:} \bar{h}_{\rm s}$ and dewetting times $t$ at $T=140\celsius C $. Rim profiles are chosen at times $t$ where  $\max_x h_\ell\approx 2\,\bar{h}_\ell$ in a),c),d) and $\max_x h_\ell\approx 3\,\bar{h}_\ell$ in b). Experimental profiles are averaged over 30 scan lines of a straight front. a),b) $\bar{h}_\ell {:} \bar{h}_{\rm s} \approx 1{:}1 = (248\pm2)\,{\rm nm} {:} (256\pm2)\,{\rm nm}$, $t=32$\,h (a) and $t=103$\,h (b). c) $\bar{h}_\ell {:} \bar{h}_{\rm s} \approx 1{:}2 = (47\pm1)\,{\rm nm} {:} (90\pm2)\,{\rm nm}$, $t=4$\,h
  d) $\bar{h}_\ell {:} \bar{h}_{\rm s} \approx 2{:}1 = (89\pm2)\,{\rm nm} {:} (44\pm2)\,{\rm nm}$, $t=16$\,h.}  
 	\label{fig:profiles1}
\end{figure*}
\noindent
To prepare those samples we used standard thin film preparation techniques as described in \cite{bommer2013droplets}.
The typical film-thicknesses $\bar{h}_\ell,\bar{h}_{\rm s}$  used in our dewetting experiments range from $45\,{\rm nm}$ to $250\,{\rm nm}$ and we performed experiments for various ratios
$\bar{h}_\ell{:}\bar{h}_{\rm s}$. The dewetting experiments were conducted at a temperature of $T=140\,{\celsius}C$. The shape of the PS-air and PMMA-air interface can be determined \emph{in situ} using AFM. Quenching the sample to room temperature the shape of the buried PS-PMMA interface can be additionally determined by AFM after stripping the upper PS layer with a selective solvent (cyclohexane, Sigma Aldrich). The full shape of all polymer interfaces is obtained by composing PS-air, PMMA-air and PS-PMMA surfaces, a procedure\footnote{The composition of the 3-$D$ image requires rotation, shift and tilt of upper and lower AFM scan as postprocessing for a perfect match. Cross sections are averaged over a few scan lines.} which generates shapes as shown in  FIG.~\ref{fig:AFM3D}.
Both polymers were purchased from Polymer Standard Service Mainz (PSS-Mainz,Germany) with  polydispersity of $M_w/M_n=1.05$ and molecular weights of $M_w=64\,{\rm kg}/{\rm mol}$ and $M_w=9.9\,{\rm kg/mol}$ for PS and PMMA, respectively. The viscosities were measured using the self-similarity in stepped polymer films as presented in \cite{Gra} and
\cite{sal}, resulting in viscosities of PS $\mu_\ell = 700\,{\rm kPa\,s}$ and PMMA $\mu_{\rm s}=700\,{\rm kPa\,s}$. While errors of most experimental parameters lie within a few percent, the experimental errors of the viscosity values are within several ten percent being the main source of uncertainty when comparing experimental and simulation timescales.
Following the procedure described in \cite{bommer2013droplets} we experimentally determined the involved surface tensions to $\gamma_{\ell}=32.3\,{\rm mN/m}$, $\gamma_{\rm s}=32\,{\rm mN/m}$ and $\gamma_{\rm s,\ell}=1.2 \pm 0.1\,{\rm mN/m}$, compatible with values reported in literature, \emph{e.g.} in \cite{wu1970surface}. The parameters $\mu_\text{s}:\mu_\ell\sim 1$ and $\bar{h}_s:\bar{h}_\ell\sim 1$ suggest, that initially dissipation in the substrate and in the liquid are of the same order.

The flow describing the dewetting process of highly viscous Newtonian fluids is modeled by the Stokes equation
\begin{subequations}
\begin{align}
 - \nabla\cdot\boldsymbol{\tau}_i+\nabla p_i = \mathbf{f},\\
\nabla\cdot\mathbf{u}_i=0,
\end{align}
\end{subequations}
with shear stress $\boldsymbol{\tau}_i=\mu_i(\nabla \mathbf{u}_i+\nabla \mathbf{u}_i^T)\bigr)$ and
solved in $\Omega_i$ for $i\in\{{\rm s},\ell\}$ in the substrate and the liquid. Assuming translational invariance of the 3-$D$ flow in $y$-direction, one can parameterize domains and boundaries
at time $t$
\begin{subequations}
\begin{align}
&\Omega_{\rm s}(t)=\{(x,z)\in\mathbb{R}^{3}:0<z<h_{\rm s}\},\\
&\Omega_{\ell}(t)=\{(x,z)\in\mathbb{R}^{3}:h_{\rm s}<z<h_{\rm s}+h_\ell\},\\
&\Gamma_{\rm s}(t)=\{(x,z)\in\mathbb{R}^{3}:(z=h_{\rm s})\cap (h_\ell=0)\},\\
&\Gamma_{\ell}(t)=\{(x,z)\in\mathbb{R}^{3}:(z=h_{\rm s}+h_\ell)\cap (h_\ell>0)\},\\
&\Gamma_{\rm s,\ell}(t)=\{(x,z)\in\mathbb{R}^{3}:(z=h_{\rm s})\cap (h_\ell>0)\},
\end{align}
\end{subequations}
using non-negative functions $h_{\rm s}(t,x),h_\ell(t,x)$ as shown in FIG.~\ref{fig:AFM3D}. The flow field, pressure, and viscosity in $\Omega_i$ are denoted by $\mathbf{u}_i,p_{i},\mu_i$.
The equations in the two regions are coupled by interface/boundary conditions,  no-slip $\mathbf{u}_{\rm s}=0$ at $z=0$, continuity $\mathbf{u}_\ell=\mathbf{u}_{\rm s}$ on $\Gamma_{s,\ell}$,  tangential and normal stress conditions due to surface tension at the free surfaces $\Gamma_s,\Gamma_\ell$, and corresponding jump conditions on the interface $\Gamma_{s,\ell}$.
Kinematic conditions relate the flow field to the velocities of surfaces and interfaces.
At the contact line
one imposes further conditions on the triple junction using the Neumann triangle.
The equations are rescaled using
\begin{align}
 \label{eqn:scaling}
[x]=[z]\, =\, H,\qquad [t]\,=\, H {\mu_\ell}/{\gamma_{\rm s}}, \qquad
\end{align}
and replace the dimensional parameters by $\tilde{\gamma}_{\rm s}=1$, $\tilde{\gamma}_{\ell}=\gamma_{\ell}/\gamma_{\rm s}$, $\tilde{\gamma}_{{\rm s},\ell}=\gamma_{{\rm s},\ell}/\gamma_{\rm s}$, and $\tilde{S}=S/\gamma_{\rm s}$.
Consequently, all experimental and numerical lengths are normalized to the initial film height $H=\bar{h}_\ell$ and times are rescaled using $t/\bar{h}_\ell$.
Following the standard thin-film approximation we assume that the interfaces are
shallow $|\partial_x h_{\rm s}|\ll 1$, $|\partial_x h_\ell|\ll 1$. Then, a formal asymptotic calculation shows that  $h_{\rm s},h_\ell$ are solutions of a system of degenerate parabolic equations %
defined separately on the wetted region $\omega(t)=\{x\in\mathbb{R}:h_\ell(t,x)>0\}$
and its complement.
 For $x\in\omega(t)$ we have
\begin{subequations}
\label{eqn:tf}
\begin{align}
\partial_t h_{\rm s} &= \partial_x \left( M_{11} \partial_x \pi_{1} + M_{12}\partial_x\pi_2\right),\label{prec1}\\
\partial_t h_\ell   &= \partial_x \left( M_{21} \partial_x \pi_{1} + M_{22}\partial_x\pi_2\right),\label{prec2}
\end{align}
with mobilities $M_{11}=h_{\rm s}^3/(3\mu)$, $M_{12}{=}M_{21}{=}h_{\rm s}^2h_\ell/(2\mu)$, $M_{22}{=}h_\ell^3/3 + h_{\rm s} h_\ell^2/\mu$,
pressures
$\pi_{1}{=}-(\tilde{\gamma}_{\rm s,\ell}+\tilde{\gamma}_{\ell})\partial_{xx} h_{\rm s} - \tilde{\gamma}_{\ell}\partial_{xx} h_\ell$,
$\pi_2{=}-\tilde{\gamma}_{\ell}\partial_{xx} h_{\rm s} - \tilde{\gamma}_{\ell}\partial_{xx} h_\ell$, and %
viscosity ratio $\mu=\mu_{\rm s}/\mu_\ell$. On the complement of $\omega$,
only $h_{\rm s}$ is unknown and solves
$\partial_t h_{\rm s} = \partial_x\left(m\,\partial_{x} \pi_{1} \right)$
\end{subequations}
with $m=h_{\rm s}^{3}/(3\mu)$ and pressure $\pi_{1}=-\partial_{xx} h_{\rm s}$. Additional boundary conditions and kinematic conditions are imposed at the contact line $x_{\rm c}=\partial\omega$.
Note that once the solution is known, then the thin-film approximation returns the horizontal component of the flow field $u_{i}=\mathbf{u}_{i}\cdot\hat{\mathbf{x}}$ using
\begin{subequations}
 \label{eqn:velreconstruct}
\begin{align}
u_{\rm s}&=\frac{\partial_{x}\pi_{1}}{2\mu}z^{2}+c_{{\rm s},1}z+c_{{\rm s},2},\\
u_{\ell}&=\frac{\partial_{x}\pi_{2}}{2}z^{2}+c_{\ell,1}z+c_{\ell,2},
\end{align}
\end{subequations}
for $x\in\omega$. The functions $c_{{\rm s},1},c_{{\rm s},2},c_{\ell,1},c_{\ell,2}$ depend on $(t,x)$ and are determined using the boundary conditions $u_{\rm s}=0$ at $z=0$, $u_{\rm s}-u_\ell=\partial_z(u_{\rm s}-\mu u_\ell)=0$ at $z=h_{\rm s}$,
$\partial_z u_\ell=0$ at $z=h_{\rm s}+h_\ell$ as before. The flow field in the complement is determined analogously.
The formal derivation of this model can be found in \cite{kriegsmann2003steady}. The corresponding novel numerical scheme, which in particular uses the energetic variational structure to discretize the Neumann triangle, is described in \cite{hjkp14}.
%
%
\section{Discussion of shapes and rates}
\noindent
Using identical rim heights to compare experimentally measured and theoretically computed interface profiles, we find an excellent agreement of both the characteristic shapes of the liquid-air and of the liquid-liquid interfaces, FIGS.~\ref{fig:profiles1}a)-d). The material of the dewetting liquid (PS) accumulates in a rim which, by conservation of mass, grows in time when the liquid retracts from the substrate (PMMA). Away from the rim the interfaces decay in an oscillatory fashion into their prepared constant states $h_{\rm s}(t,x),h_\ell(t,x)\to \bar{h}_{\rm s},\bar{h}_\ell$. Also some material of the substrate is dragged along generating a depletion near $x<x_{\rm c}$ and an accumulation of substrate material near $x>x_{\rm c}$. Right next to the contact line, the liquid-liquid interface extends deeply into the substrate and generates a trench and thereby produces additional resistance against the dewetting motion. The size of this trench depends only weakly on the size of the dewetting rim, \emph{i.e.}, the dewetting distance. The contact line itself is elevated by the flow, a dynamic feature not observed in stationary droplets for sufficiently thick substrates \cite{bommer2013droplets} but quite common for soft substrates, \emph{e.g.},  see \cite{style2013universal}.

\begin{figure}[h]
\centering
\includegraphics[width=0.6\linewidth]{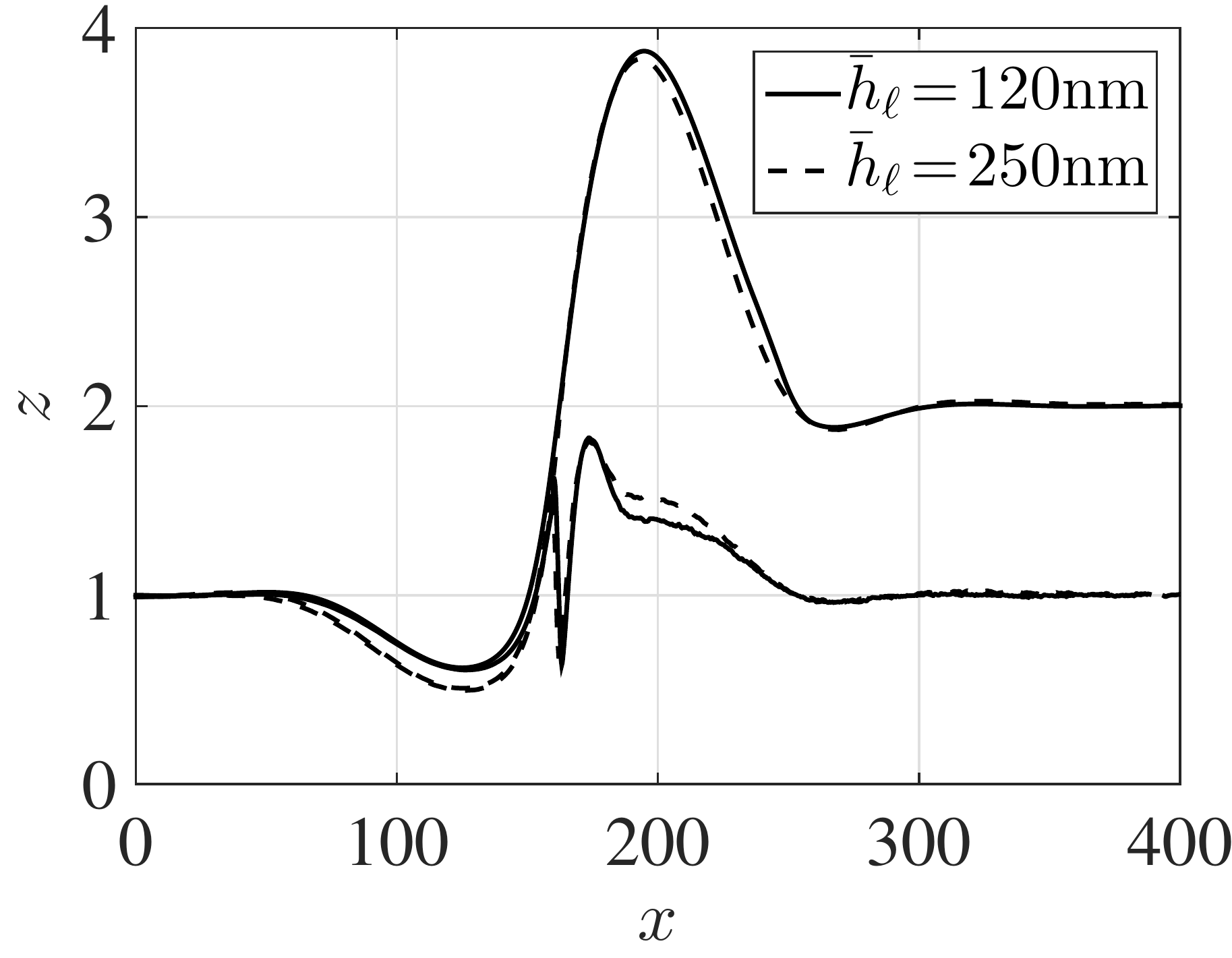}%
\caption{Experiments with film thickness ratio $\bar{h}_\ell{:}\bar{h}_{\rm s}\,\approx\,1{:}1$ and different absolute film thicknesses when the rim height is about $\max_x h_\ell \approx 3\bar{h}_\ell$. Dashed lines: $\bar{h}_\ell = (248 \pm 2)\,{\rm nm}{:}\bar{h}_{\rm s} = (256 \pm 2)\,{\rm nm}$, solid line:  $\bar{h}_\ell = (117 \pm 3)\,{\rm nm}{:}\bar{h}_{\rm s} = (122 \pm 2)\,{\rm nm}$.}
\label{fig:profiles2}
\end{figure}

\begin{figure*}[ht]
\centering
\includegraphics[width=0.35\linewidth]{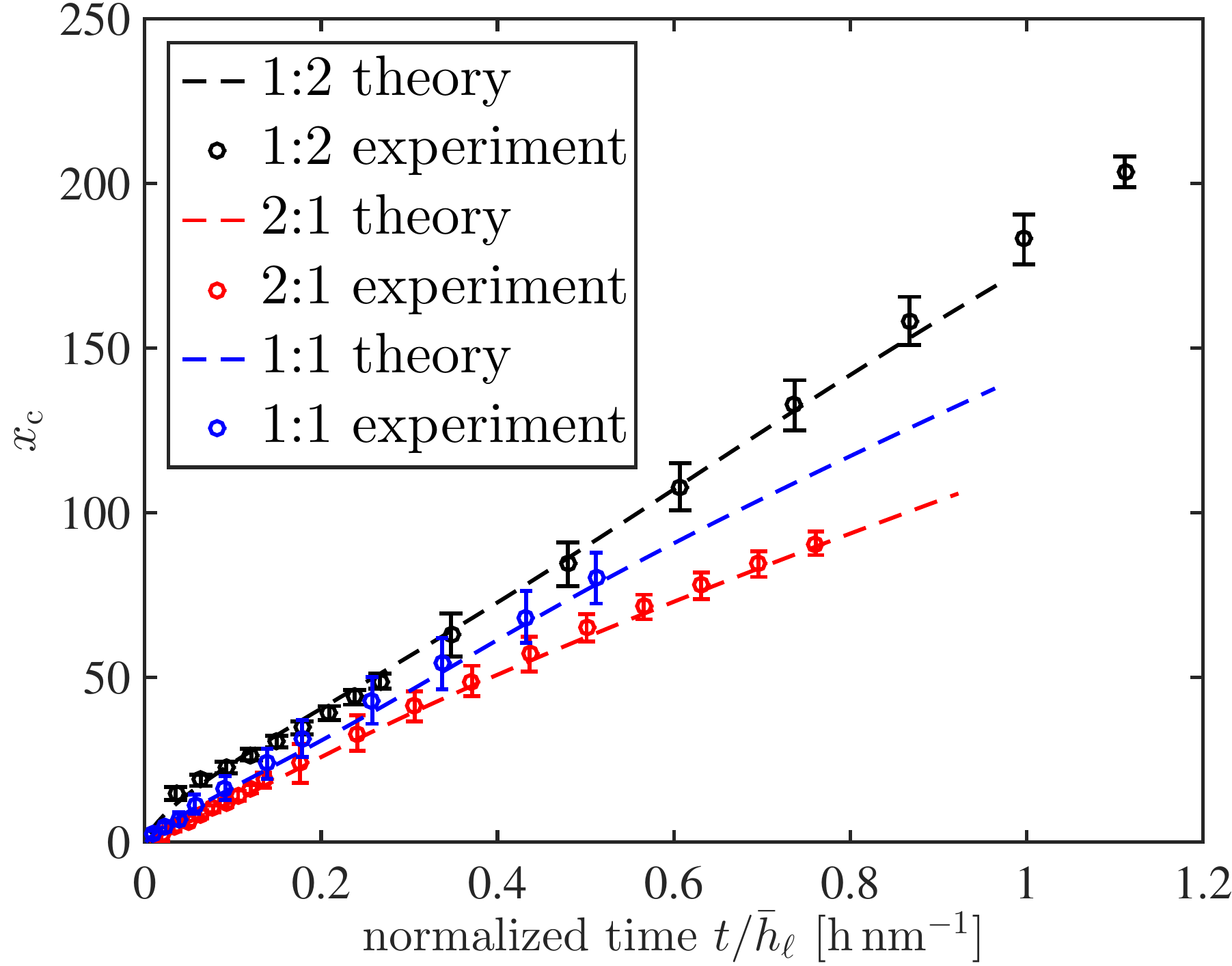}%
\qquad\qquad%
\includegraphics[width=0.35\linewidth]{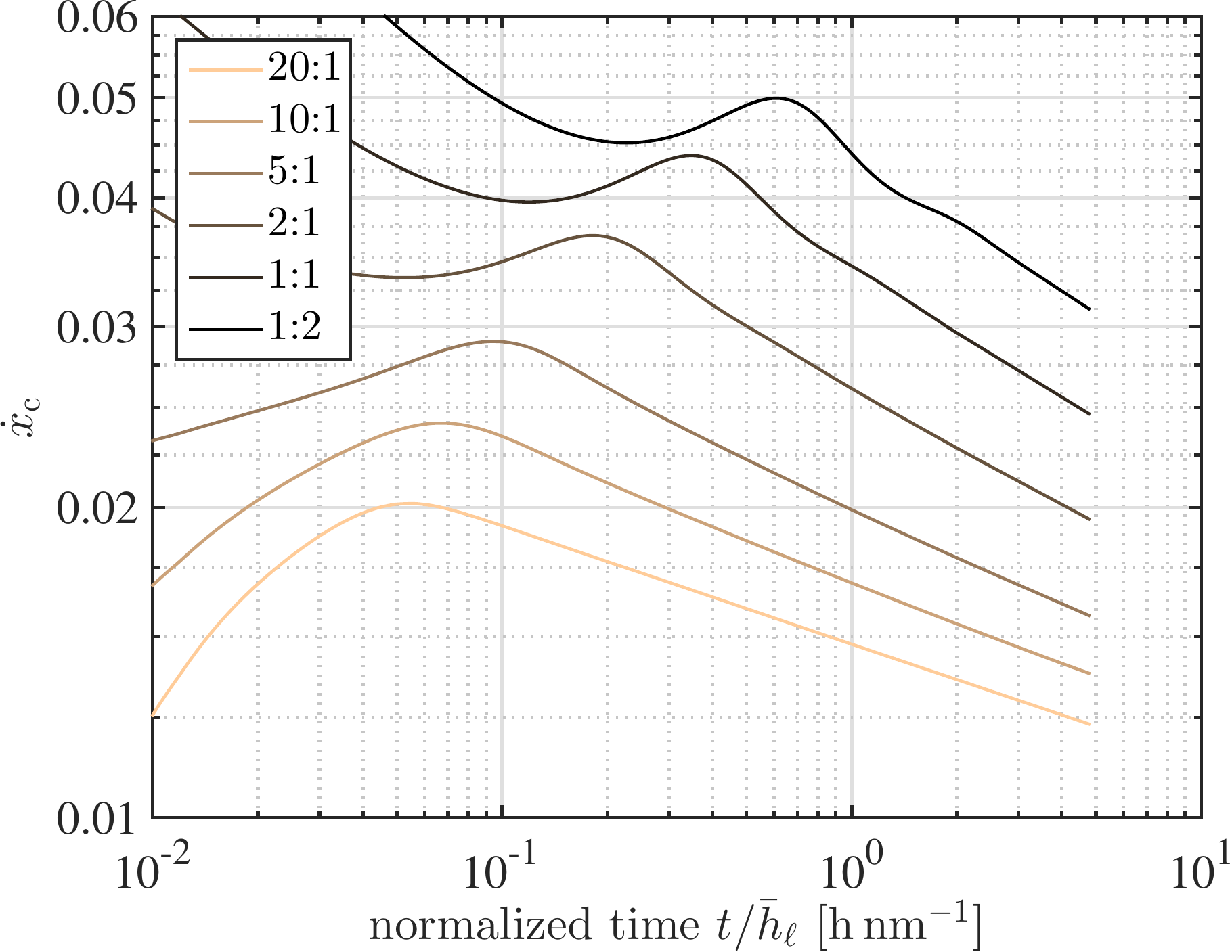}%
\caption{(left) Non-dimensional dewetted distance $x_{\rm c}$ for thickness ratios
$\bar{h}_\ell {:} \bar{h}_{\rm s}\,\approx\,1{:}1 = (248\pm2)\,{\rm nm}{:}(256\pm2)\,{\rm nm}$,
$\bar{h}_\ell {:} \bar{h}_{\rm s}\,\approx\,2{:}1 = (89\pm2)\,{\rm nm}  {:}(44\pm2)\,{\rm nm}$,
$\bar{h}_\ell {:} \bar{h}_{\rm s}\,\approx\,1{:}2 =(47\pm1)\,{\rm nm} {:} (90\pm2)\,{\rm nm}$
from experiment and simulation suggesting constant dewetting rates
$\dot{x}_{{\rm c},1{:}1}\approx 4.4\!\cdot\!10^{-2} {\rm nm/s}$,
$\dot{x}_{{\rm c},2{:}1}\approx 3.0\!\cdot\!10^{-2} {\rm nm/s}$,
$\dot{x}_{{\rm c},1{:}2}\approx 5.2\!\cdot\!10^{-2} {\rm nm/s}$,
where  longer simulation times (right) prove that rates $\dot{x}_{\rm c}$ decrease depending on details such as the aspect ratio.}
\label{fig:shar}
\end{figure*}

For a fixed thickness ratio \eqref{eqn:scaling} shows that, due to the absence of other intrinsic time and length scales for Newtonian liquids, the influence of the absolute height is to scale time linearly without changing
the rescaled rim profiles. To check this prediction two liquid-liquid systems with thickness ratio $\bar{h}_\ell{:}\bar{h}_{\rm s}$ = $1{:}1$ but film thicknesses $\bar{h}_\ell\approx 100\,{\rm nm}$ and $\bar{h}_\ell \approx 250\,{\rm nm}$ were determined. An overlap of the emerged rim profiles is shown in FIG.~\ref{fig:profiles2} for corresponding dewetting distances. The good reproducibility of the characteristic rim profiles within experimental errors confirms the previously made assumption\footnote{The dewetting rates shown in FIG.~\ref{fig:shar} can be used to estimate the shear stress in the PS film to $\dot{\gamma}\approx 3\cdot10^{-4}$\,s$^{-1}$. Using the fitted polymer viscosities and the shear modulus of $G_{\rm PS}=0.2$\,MPa for PS \cite{hir} and $G_{\rm PMMA}=3$\,MPa for PMMA \cite{lomellini1992effect},
the relaxation time of the polymers $\tau_{\rm PS}\,=\,\mu_{\rm PS}/G_{\rm PS}\approx0.4$\,s and $\tau_{\rm PMMA}\approx6$\,s can be calculated. The resulting in Weissenberg numbers $\text{Wi}=\tau\, \dot{\gamma}$ are
$\text{Wi}_{\rm PMMA}\approx 6\cdot10^{-5}\ll1$ and
$\text{Wi}_{\rm PS}\approx 10^{-3}\ll1$,
so that viscoelastic effects are negligible \cite{morozov2007introductory}.}.
Compared to the ratio $\bar{h}_\ell{:}\bar{h}_{\rm s}$\,=\,$1{:}1$ in FIG.~\ref{fig:profiles1}a,b), thickness ratios of $2{:}1$ or $1{:}2$ do not lead to qualitatively new features.
For smaller aspect ratio $h_\ell{:}h_{\rm s}$\,=\,$1{:}2$, cf. FIG.~\ref{fig:profiles1}c), the above described characteristic features of the rim profile grow and for bigger aspect ratio $h_\ell{:}h_{\rm s}$\,=\,$2{:}1$, cf. FIG.~\ref{fig:profiles1}d), these features shrink in size slightly.
For $\bar{h}_{\rm s} \to 0$ we expect to observe shapes similar to that of a film dewetting from a solid substrate. The match of experiment and simulation is in all cases almost perfect, within the limits of reproducibility observable in  FIG.~\ref{fig:profiles2}.
%

A quantitative agreement between the temporal evolution of experimental and simulation rim profiles and the contact line dynamics is achieved by setting $\mu_\ell = \mu_{\rm s} =1100\,{\rm kPa\,s}$, instead of the experimentally determined value of $700\,{\rm kPa\,s}$,
for all film thicknesses and thickness ratios within the experimental accuracy, cf. Fig.~\ref{fig:shar} (left).
For small dewetting distances, the dewetting rates suggest a linear behavior $x_{\rm c}\sim t$ for all thickness ratios. For fixed substrate film thickness $\bar{h}_{\rm s}$, the dewetting rates are larger for liquid layers thinner than the substrate, $\bar{h}_\ell < \bar{h}_{\rm s}$, and smaller for thicker liquid layers, $\bar{h}_\ell > \bar{h}_{\rm s}$.
%
%
But, a close inspection of the seemingly constant dewetting rates in FIG.~\ref{fig:shar} (left) indicates that the dewetting velocity slowly decreases over time. This fact is most apparent for aspect ratio $2{:}1$, while for aspect ratio $1{:}2$ the velocity even appears to increase. However, the experimental accuracy is not sufficient to fully clarify this claim.

To clarify the dependence of the dewetting rates, results from simulation are plotted in FIG.~\ref{fig:shar} (right) for dewetting distances, which are not accessible experimentally together with further results for other film thickness ratios.
Notice the small variation in the velocities during dewetting, which explains why dewetting rates appear almost constant. However, the intricate transient behavior of the velocity $\dot{x}_{\rm c}$ featuring inflection points in the simulations coincides with the before mentioned experimental observation. For instance, for an aspect ratio of $2{:}1$ and the experimentally accessible times $t=10^{-1}\ldots\,10^0$, cf. FIG.~\ref{fig:shar} (left),  the dewetting rate decreases, while for an aspect ratio of $1{:}2$ the rate slightly increases within the observed dewetting interval. For all simulated parameters we find that for large times the velocity slowly decays to zero. 

\begin{figure*}[ht]
\centering
\includegraphics[width=0.43\textwidth]{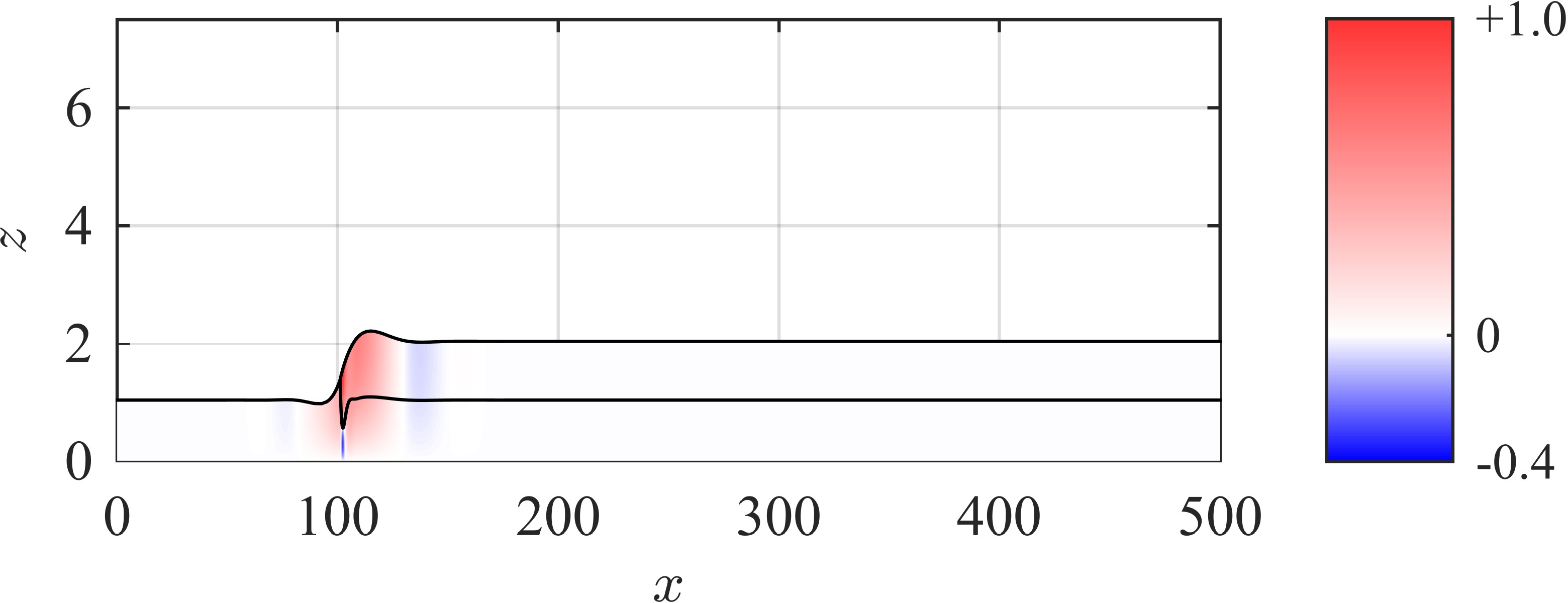}\qquad\includegraphics[width=0.43\textwidth]{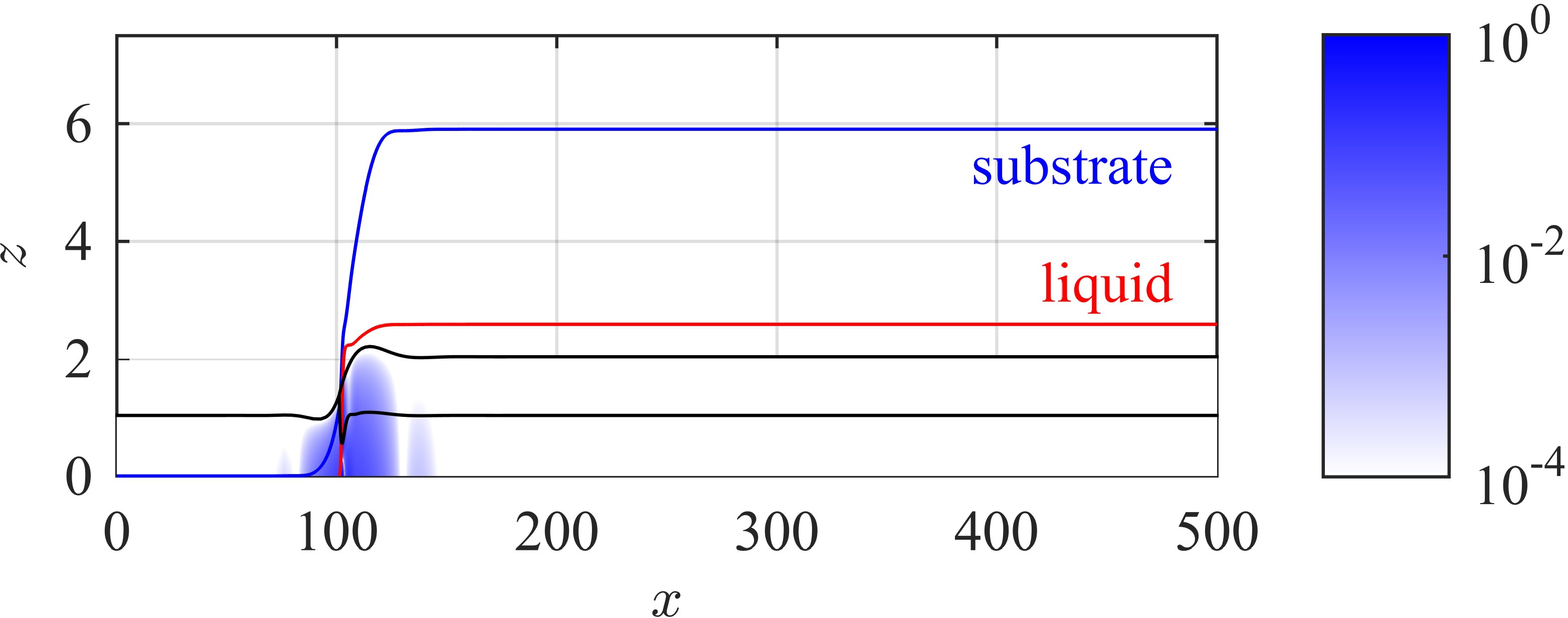}

\includegraphics[width=0.43\textwidth]{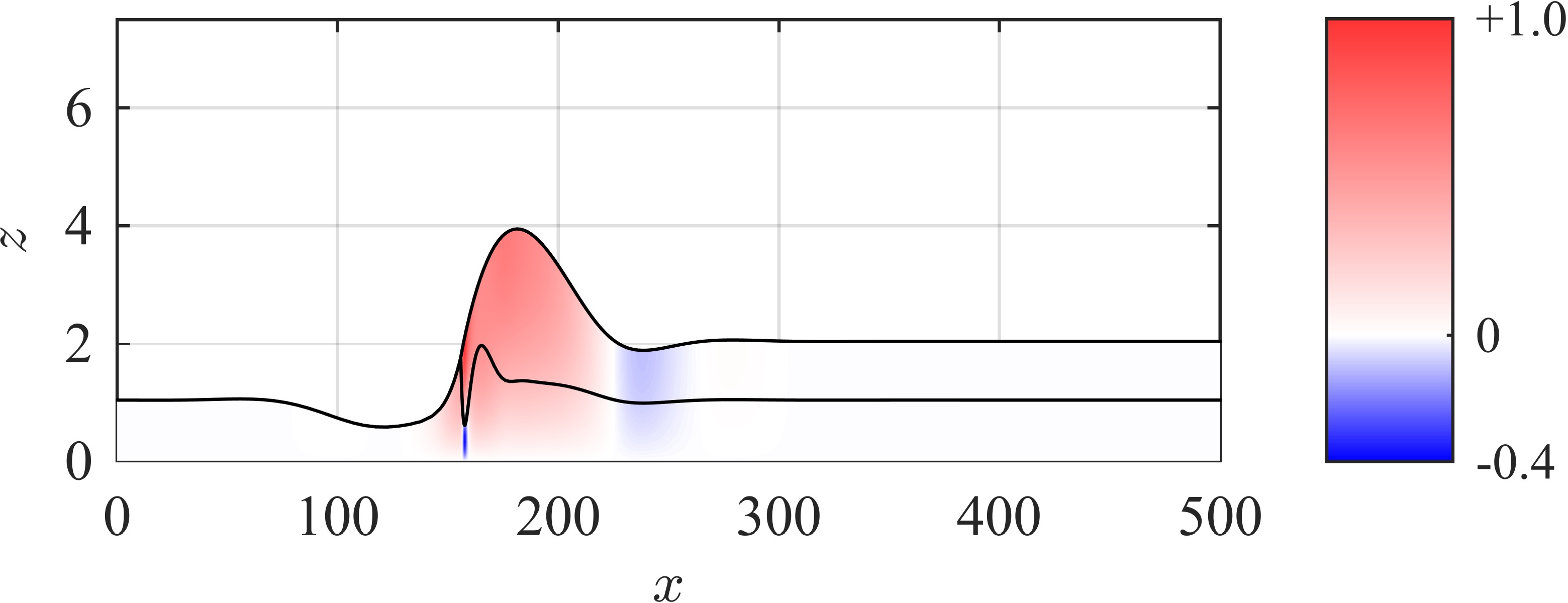}\qquad\includegraphics[width=0.43\textwidth]{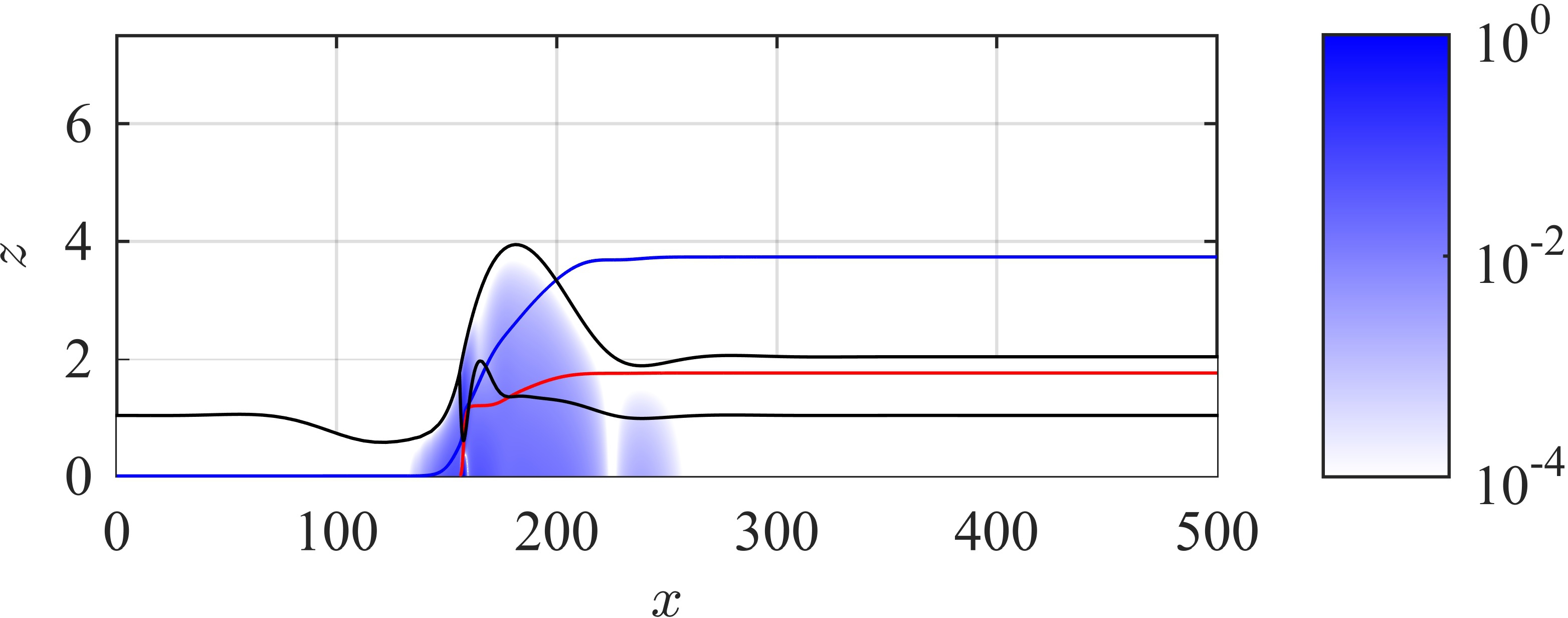}

\includegraphics[width=0.43\textwidth]{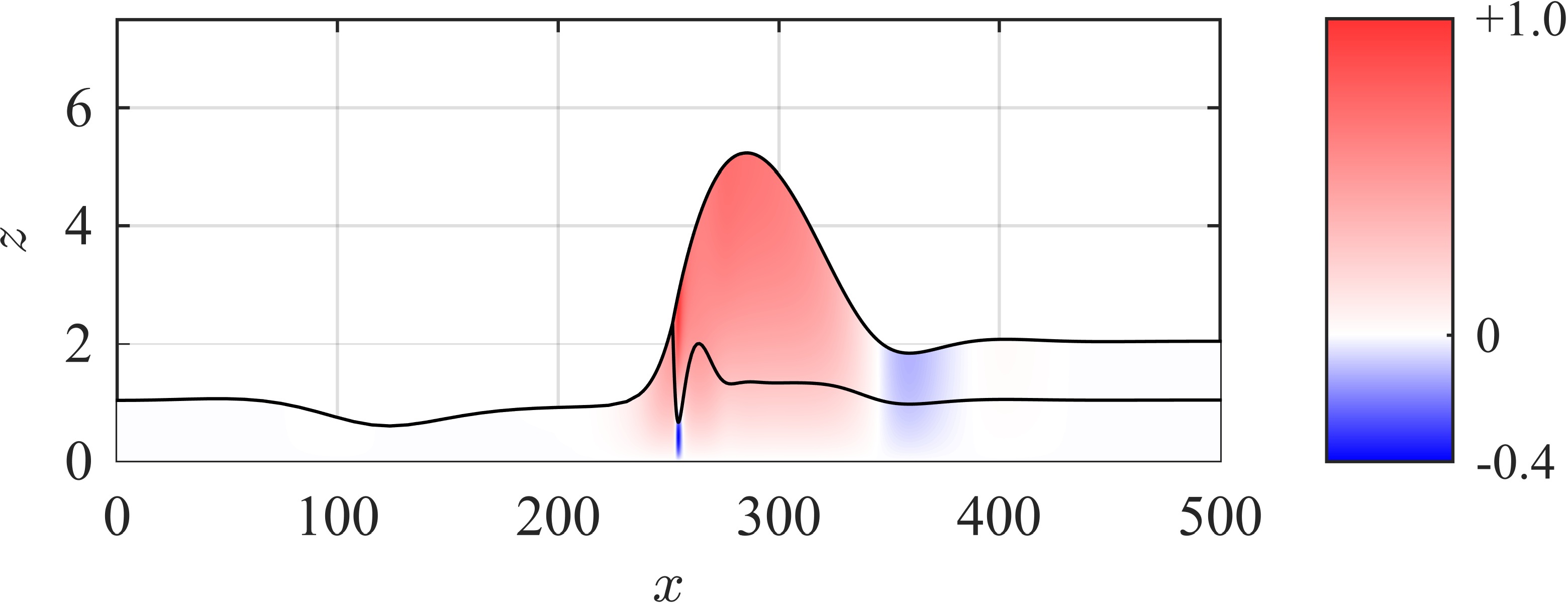}\qquad\includegraphics[width=0.43\textwidth]{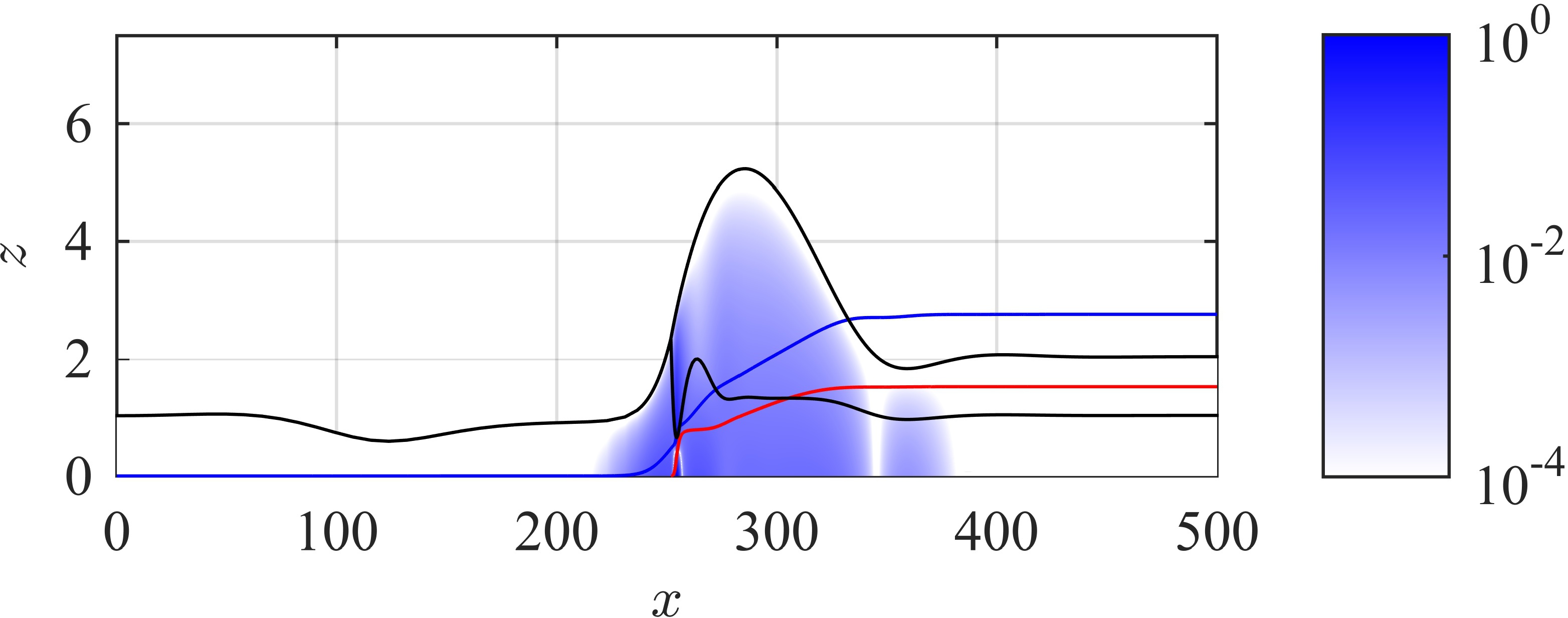}

\includegraphics[width=0.43\textwidth]{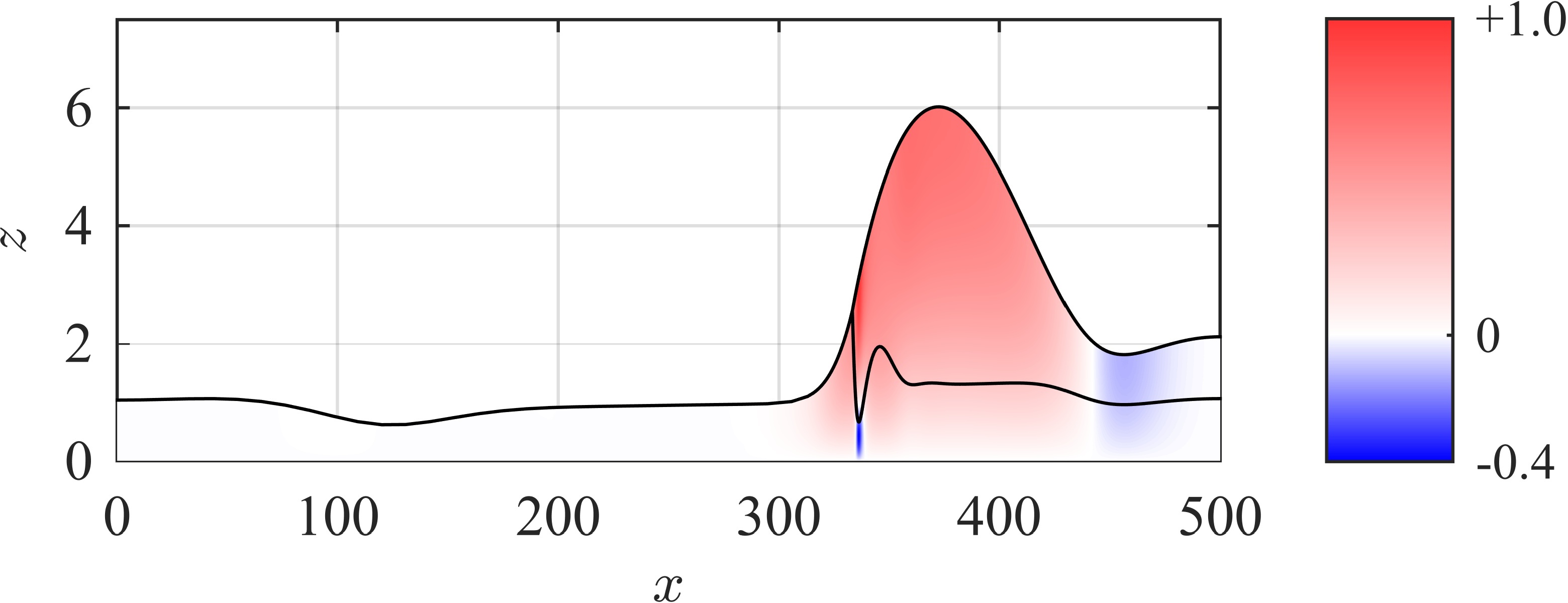}\qquad\includegraphics[width=0.43\textwidth]{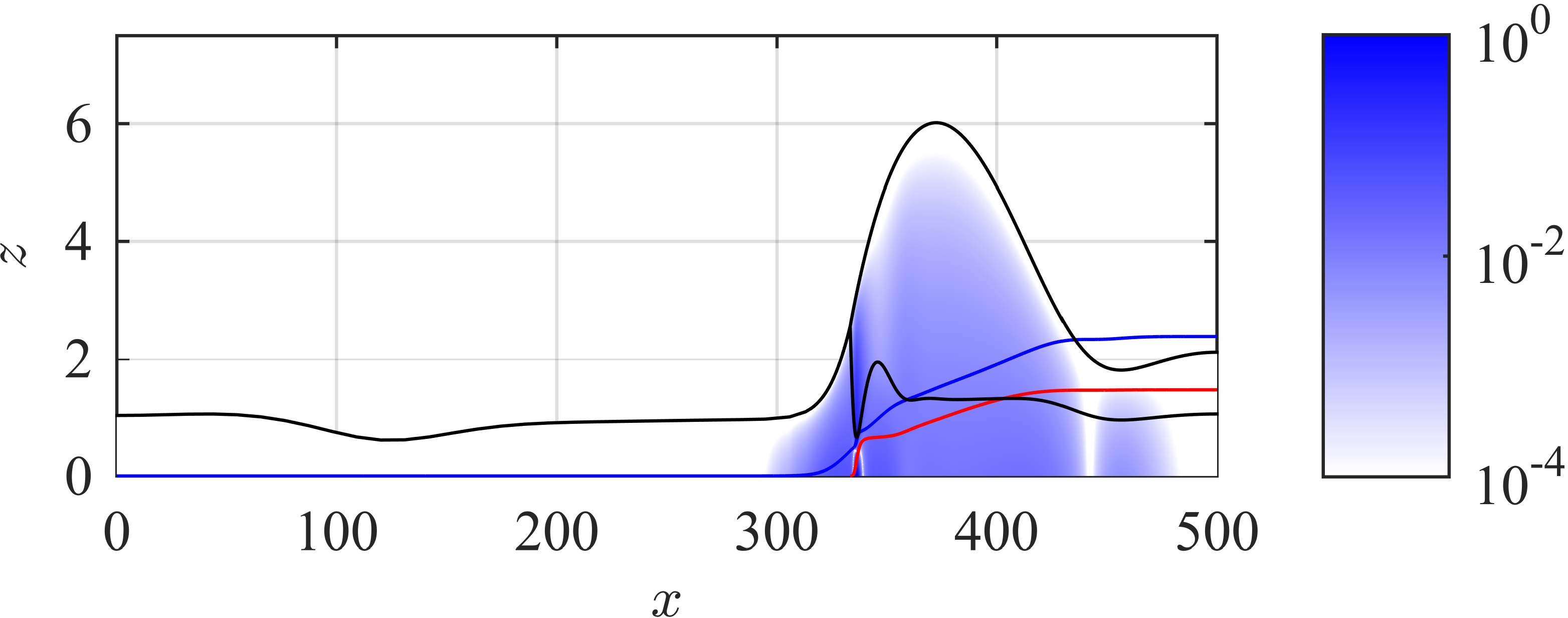}
\caption{(left) Flow fields in the substrate $u_{{\rm s}}(x,z)$ and in the liquid $u_{\ell}(x,z)$ and (right) the corresponding energy dissipation
$D_{\rm s}(x,z)=\bigl(\partial_z u_{\rm s}(x,z)\bigr)^2$ and $D_\ell(x,z)=\mu\bigl(\partial_z u_\ell(x,z)\bigr)^2$ on logarithmic scale normalized to their respective maximal values computed numerically from the thin-film model during dewetting with aspect ratio $1{:}1$ at different times increasing from top to bottom.
The additional curves show the cumulative dissipation in the substrate $\int_{-\infty}^x\int_0^{h_{\rm s}} D_{\rm s} {\rm d}z\,{\rm d}x$ (blue line)
and in the liquid $\int_{x_{\rm c}}^x\int_{h_{\rm s}}^{h_{\rm s}+h} D_\ell {\rm d}z\,{\rm d}x$ (red line) normalized with an arbitrary but time-independent constant.}
\label{fig:suppl_diss}
\end{figure*}

\section{The role of dissipation}
\noindent
Considering the dissipation balance in a 2-$D$ cross section
\begin{equation}
\label{eqn:dissi}
\mathcal{D}=\int_{\Omega_\text{s}} \tfrac{1}{2}(\partial_z u_{\rm s})^2{\rm d}\Omega + \int_{\Omega_\ell} \tfrac{\mu}{2}(\partial_z u_\ell)^2 {\rm d}\Omega = -\tilde{S}\times\dot{x}_{\rm c},
\end{equation}
the predicted slowdown of the dewetting velocity is expected since the dissipated energy also grows with rim-size.
The exact slopes in the log-log plot of FIG.~\ref{fig:shar} (right) depend on details such as thickness ratio and viscosities, and thereby do not support a universal power-law behavior. This observation confirms previous speculations by Krausch et al.~\cite{lam} about the transient nature of the experimentally measured dewetting dynamics.

Since the thin-film model accurately predicts shapes and speeds of the liquid-liquid dewetting, we extend this approach and discuss local flow features that otherwise would be inaccessible to explain the observed dewetting dynamics.
For instance, in \eqref{eqn:dissi} we see that dissipation balances with the driving surface tension
with spreading coefficient $\tilde{S}=\tilde{\gamma}_{\rm s}-(\tilde{\gamma}_{\rm s,\ell}+\tilde{\gamma}_\ell)$.
While the driving force is straightforward to understand, the dissipation depends on local details of the flow field and shows where friction is generated.
As a representative example we use a layer thickness ratio $1{:}1$ and show rim profiles at different times overlayed with the horizontal component of the velocity in FIG.~\ref{fig:suppl_diss} (left) and the corresponding dissipation $(\partial_z u_\ell)^2$ and $\mu(\partial_z u_{\rm s})^2$ (right).
The flow fields in the left panels of FIG.~\ref{fig:suppl_diss} point mainly in the positive $x$-direction with its maximum at the contact line. Away from the rim $|x-x_{\rm c}|\gg 0$ the flow field vanishes. Below the depression of the liquid-substrate interface there is  a rather strong and localized backflow in the liquid substrate. This backflow is due to conservation of mass balancing the forward transport of the depression.

Due to the boundary conditions $\partial_z u_{{\rm s},\ell}=0$ the dissipation vanishes at the liquid/air and substrate/air interfaces, whereas the flow field is zero at the solid/substrate interface $z=0$, resulting in a large shear rate and a large energy dissipation, cf. right panels of FIG.~\ref{fig:suppl_diss}.
Close to the backflow and close to the contact line the maximal dissipation density is reached.

%
%
%
However, due to the small size of these regions the integrated dissipation near the contact line and in the remaining rim are of the same order, at least for the transient times and moderately large rims considered here.
To emphasize this fact, we additionally included the cumulative dissipation in FIG.~\ref{fig:suppl_diss} for different times.
Since the shear rate is large at the solid interface where $z=0$, clearly the dissipation for an aspect ratio $1{:}1$ is large in the substrate for the short and intermediate times considered experimentally. Nevertheless, with the volume of the liquid rim increasing in time, ultimately the dissipation in the liquid layer will dominate for large times or for higher aspect ratios. Similarly, one can identify two different zones where the energy is dissipated.
Specifically shown in right panels of Fig.~\ref{fig:suppl_diss}, a significant amount
of the cumulative dissipation is produced in a small region near the contact line. 

 This is visible in the  steep increase in the cumulative dissipation in that area. The remaining contribution to the dissipation is more or less evenly distributed over the rim width resulting in a moderate and constant increase of the cumulative dissipation over the width of the rim. For large times this bulk contribution will dominate the dissipation and forces the velocity to decay to zero. This can also be seen in the temporal evolution of the dissipation profiles.

The variable contributions to the energy dissipation impacts the general derivation of power-law dewetting rates. This impact can be explained on the basis of the known dewetting behavior on a solid substrate: In the intermediate slip model \cite{fetzer2005new}, the dominant contribution comes from a substrate dissipation, so that the total dissipation is proportional to the rim width and the squared dewetting velocity leading to a $x_\text{c}\sim t^{2/3}$ dewetting law. Another example from the above reference is the no-slip model, where the dissipation is localized near the contact line and thereby one obtains a linear dewetting law $x_\text{c}\sim t$ with logarithmic corrections.
Accordingly, the power-law dewetting rates predicted by Brochard et al.~\cite{bro} rely on the assumption that the dissipation is generated in only one such localized zone together with a nearly self-similar growth of rim shapes.
The failure of this assumption explains why in our setting
the liquid-liquid dewetting process is not in a regime that admits a simplification to a power-law rate, and thereby challenges the applicability of the theoretical results in \cite{joanny1987wetting,bro} to experimental systems.
In contrast, the consideration of the liquid-liquid dewetting using thin film models with explicit contact line dynamics, conducted here, allows to describe the variable energy dissipation in a liquid-liquid system and to quantitatively derive rim shapes and dewetting rates.

\section{Conclusion}
\noindent
Motivated by the
long standing discrepancy between theoretically predicted and experimentally observed rates
for liquid-liquid dewetting, we performed a combined theoretical and experimental investigation of the transient interface shapes and dewetting rates. Conducting a full simulation of the sharp interface thin-film model for Newtonian liquids without any a priory assumptions on rim shape development or energy dissipation we obtained a full agreement with experimentally determined interface shapes and dewetting dynamics using the relevant experimental parameters like viscosities and surface thicknesses.
As the main tool to assess the transient nature of the flow, we reconstructed local flow and dissipation fields. Such a detailed analysis of a local energy balance provides deep insights into underlying mechanisms driving a process.

By analyzing the local energy dissipation, we have found that the liquid-liquid dewetting system is in a transient state with no self-similar behavior and with no power-law dewetting rate.
A similar energetic argument provided the explanation why, for very large times beyond experimental reach, the dewetting velocity slowly decreases to zero.
Such predictions would be impossible using heuristic approaches, since the transient internal flow is rather complex and results from a complex interaction of substrate and liquid. This explains further why this theoretical study achieved satisfying agreement with experimental results and a conclusive understanding of liquid-liquid dewetting in contrast to existing theoretical considerations. The demonstrated ability to use energetic arguments to quantitatively describe liquid-liquid systems thus set grounds for a similarly complete understanding as already obtained for liquid-solid dewetting systems. It might be in particular possible to extend this approach also to fluids with complex rheological behavior.

\section*{Acknowledgements}
{This work was supported by the DFG SPP 1506 \emph{Transport processes at fluidic interfaces} under grant SE1118/5 and WA1653/4 and by the Einstein Foundation Berlin in the framework of the \textsc{Matheon} project OT1/OT8.}


\bibliographystyle{abbrvnat}
\bibliography{LLD}
\end{document}